\documentstyle[12pt,fleqn,qul]{article}

\newcommand{\anti}[1]{\overline{#1}}

\newcommand{\AmS}{{\protect\the\textfont2
  A\kern-.1667em\lower.5ex\hbox{M}\kern-.125emS}}

\title{MESON-EXCHANGE MODEL FOR THE $YN$ INTERACTION\footnote{
Presented at Sendai International Workshop on the Spectroscopy
of Hypernuclei, January 1998, Sendai, Japan}}
\author{J. Haidenbauer, W. Melnitchouk and J. Speth
\address{{\em Institut f\"ur Kernphysik,
        Forschungszentrum J\"ulich, \\ 
        D-52425 J\"ulich, Germany}}}

\begin{document}  

\maketitle

\begin{abstract}
  We report on progress in the development of a microscopic model for the
hyperon-nucleon interaction within the meson-exchange framework, which
incorporates both one-meson as well as correlated two-meson exchange
contributions.  The main new feature of the model is the exchange of two
correlated pions or kaons, both in the scalar-isoscalar and the
vector-isovector channels.  In the scalar channel it provides the main
part of the intermediate-range interaction between two baryons, thereby
eliminating the need for the fictitious sigma meson, which has been an
unavoidable element of effective one-boson-exchange descriptions.
\end{abstract}

\section{INTRODUCTION}

The role of strangeness in low and medium energy nuclear physics is currently
of considerable interest, as it has the potential to deepen our
understanding of the relevant strong interaction mechanisms in
the non-perturbative regime of QCD.
For example, the system of a strange baryon (hyperon $Y$) and a
nucleon ($N$) is in principle an ideal testing ground for investigating
the importance of SU(3)$_{flavor}$ symmetry in hadronic interactions.
Existing meson exchange models of the $YN$ force usually assume SU(3)
flavor symmetry for the hadronic coupling constants, and in some
cases \cite{Holz,Reu} even the SU(6) symmetry of the quark model.
The symmetry requirements provide relations between couplings of
mesons of a given multiplet to the baryon current, which greatly
reduce the number of free model parameters.
Specifically, coupling constants at the strange vertices are then
connected to nucleon-nucleon-meson coupling constants, which in
turn are constrained by the wealth of empirical information on $NN$
scattering.  Essentially all these $YN$ interaction models can reproduce the 
existing $YN$ scattering
data, so that at present the assumption of SU(3) symmetry for the
coupling constants cannot be ruled out by experiment.

One should note, however, that the various models differ dramatically
in the treatment of the scalar-isoscalar meson sector, which describes
the baryon-baryon interaction at intermediate ranges.
For example, the Nijmegen group \cite{NijII,NijIII,NijIV} views this
interaction as being generated by genuine scalar meson exchange.
In their model D \cite{NijII} an $\epsilon(760)$ is exchanged as an
SU(3)$_{\it flavor}$ singlet.
In models F~\cite{NijIII} and NSC~\cite{NijIV}, a scalar SU(3) nonet is
exchanged --- namely, two isospin-0 mesons (besides the $\epsilon(760)$, 
the $\epsilon '(1250)$ in model F and $S^*(975)$ in model NSC), an
isospin-1 meson $\delta$ and an isospin-1/2 strange meson $\kappa$.
The T\"ubingen model \cite{Tueb}, on the other hand, which is
essentially a constituent quark model supplemented by $\pi$ and
$\sigma$ exchange at intermediate and short ranges, treats the
$\sigma$ meson as an SU(3) singlet with a mass of 520 MeV.

\begin{figure}[h]
\vskip 4cm
\includegraphics{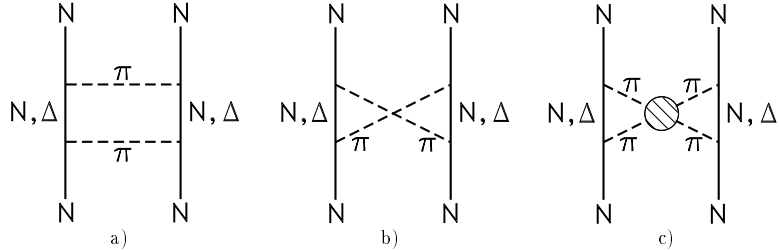} 
\caption{Two-pion exchange in the $NN$ interaction: (a) uncorrelated
	iterative and (b) crossed boxes, (c) correlated two-pion exchange.}
\label{fig1}
\end{figure}

In the (full) Bonn $NN$ potential~\cite{MHE} the intermediate range
attraction is provided by uncorrelated and correlated $\pi\pi$ exchange
processes (Figs.~\ref{fig1}(a)--(b) and Fig.~\ref{fig1}(c), respectively),
with $NN$, $N\Delta$ and $\Delta\Delta$ intermediate states.
{}From earlier studies of the $\pi\pi$ interaction it is known that
$\pi\pi$ correlations are important mainly in the scalar-isoscalar
and vector-isovector channels.
In one-boson-exchange (OBE) potentials these are included effectively
via exchange of sharp mass $\sigma$ and $\rho$ mesons.
One disadvantage of such a simplified treatment is that this
parameterization cannot be transferred into the hyperon sector
in a well defined manner.
Therefore in the earlier $YN$ interaction models of the J\"ulich 
group~\cite{Holz}, which start from the Bonn $NN$ potential,
the coupling constants of the fictitious $\sigma$ meson at the
strange vertices ($\Lambda\Lambda\sigma$, $\Sigma\Sigma\sigma$)
are free parameters --- a rather unsatisfactory feature of the
models.
This is especially true for the extension to the strangeness $S=-2$
channels, interest in which initiated with the prediction of the
H-dibaryon by Jaffe~\cite{Jaffe}.
Unfortunately, so far there is no empirical information about these
channels.

These problems can be overcome by an explicit evaluation of correlated
$\pi\pi$ exchange in the various baryon-baryon channels.
A corresponding calculation was already done for the $NN$ case
(Fig.~\ref{fig1}(c)) in Ref. \cite{Kim}.
The starting point there was a field theoretic model for both the
$N\anti{N}\to\pi\pi$ Born amplitudes and the $\pi\pi$ and $K\anti{K}$
elastic scattering~\cite{Lohse}.
With the help of unitarity and dispersion relations the amplitude
for the correlated $\pi\pi$ exchange in the $NN$ interaction was
computed, showing characteristic discrepancies with the $\sigma$
and $\rho$ exchange in the (full) Bonn potential.

In a recent paper \cite{REUBER} the J\"ulich group presented a
microscopic derivation of correlated $\pi\pi$ exchange in various
baryon-baryon ($BB'$) channels with strangeness $S=0, -1$ and $-2$.
The $K\anti{K}$ channel is treated on an equal footing with the
$\pi\pi$ channel in order to reliably determine the influence of
$K\anti{K}$ correlations in the relevant $t$-channels.
In this approach one can replace the phenomenological $\sigma$
and $\rho$ exchanges in the Bonn $NN$ \cite{MHE} and J\"ulich $YN$
\cite{Holz} models by correlated processes, and eliminate undetermined
parameters such as the $BB'\sigma$ coupling constants \cite{we}.
The resulting models thus have more predictive power and should allow
a more reliable treatment of the $S=-2$ baryon-baryon channels.

\begin{figure}[hb]
\vskip 4.3cm 
\includegraphics{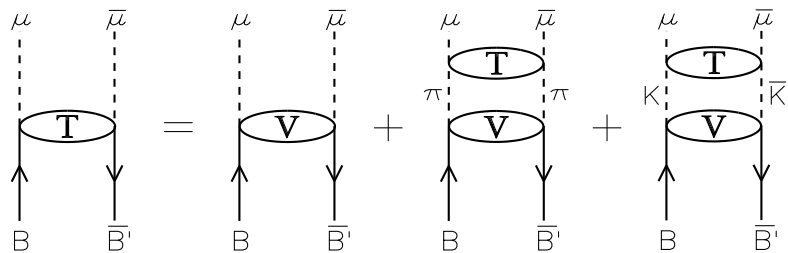} 
\caption{The present dynamical model for the
	$B\bar B \rightarrow \mu \bar \mu$
	amplitude ($\mu \bar \mu$ = $\pi\pi$, $K\bar K$).}
\label{fig5}
\end{figure} 

In the next section we describe the basic ingredients used
to derive correlated $\pi\pi$ and $K\anti{K}$ exchange potentials for
the baryon--baryon amplitudes in the $\sigma$ and $\rho$ channels.
We also give a short outline of the microscopic model for the required
$B\anti{B'}\to\pi\pi,\,K\anti{K}$ amplitudes.
The results for the potentials are presented in Section 3, where we
focus on the $\pi\pi$ correlations in the $\sigma$ channel.
These are compared with the results obtained from the exchange 
of a sharp mass $\sigma$ meson in the Bonn $NN$ \cite{MHE} and J\"ulich $YN$
\cite{Holz} potentials.
Furthermore, we introduce and discuss results obtained with a
parameterization of correlated $\pi\pi$ and $K\anti{K}$ exchange
potentials by an effective $\sigma$ exchange for the $NN$ and $YN$
channels. 
Finally, some concluding remarks are made in Section 4.

\section{MODEL FOR CORRELATED $2\pi$ EXCHANGE}

Figure \ref{fig5} shows a graphic representation of our dynamical model
for correlated $2\pi$ exchange.
Here $B\anti{B'}$ stands for $N\anti{N}$, $\Lambda \anti{ \Lambda}$,
$\Lambda \anti{ \Sigma}$/$\Sigma \anti{ \Lambda}$
or $\Sigma \anti{ \Sigma}$.
The basic ingredients are the $B\anti{B'} \rightarrow \pi\pi$,
$K\anti{K}$ Born amplitudes and the $\pi\pi$-$K\anti{K}$ interaction,
which we outline below.
Note that a microscopic model of correlated $\pi\pi$ exchange
for the $NN$ case was already presented in Ref.~\cite{Kim}.
Interestingly enough, the resulting strength turned out to be
considerably larger than that from sharp mass $\sigma '$ and $\rho$
exchanges used in the (full) Bonn potential \cite{MHE}.
 
\subsection{$\pi\pi \rightarrow \pi\pi$ Amplitude} 

\begin{figure}[t]
\vskip 4cm 
\includegraphics{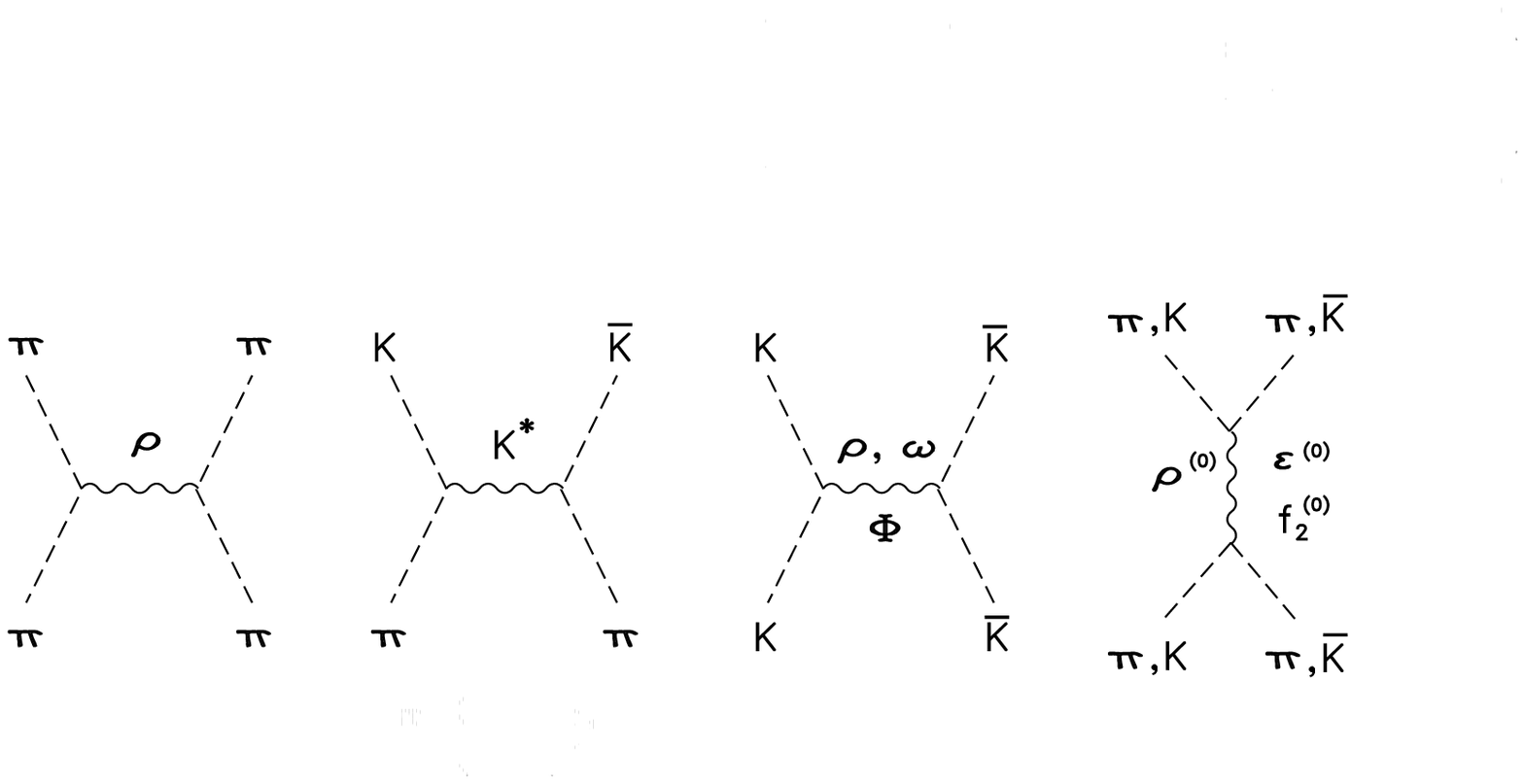} 
\caption{Meson exchange diagrams included in the dynamical model for
	the $\pi\pi$, $K\anti{K}$ interaction \protect\cite{Lohse}.}
\label{fig6}
\end{figure}

The dynamical model used here is based on the meson exchange framework
of Refs. \cite{Lohse,Janssen} involving the $\pi\pi$ and $K\anti{K}$ coupled
channels.
The driving terms for $\pi\pi \rightarrow \pi\pi$ consist of exchange
and pole diagrams (Fig.~\ref{fig6}, first and last diagram, respectively)
with $\epsilon \equiv f_0(1440)$, $\rho \equiv \rho (770)$ and
$f_2 \equiv f_2(1274)$ intermediate states.
The coupling $\pi\pi \rightarrow K\anti{K}$ is provided by $K^*(892)$
exchange, illustrated in the second diagram in Fig.~\ref{fig6}.

\begin{figure}[htb]
\vskip 12cm
\includegraphics{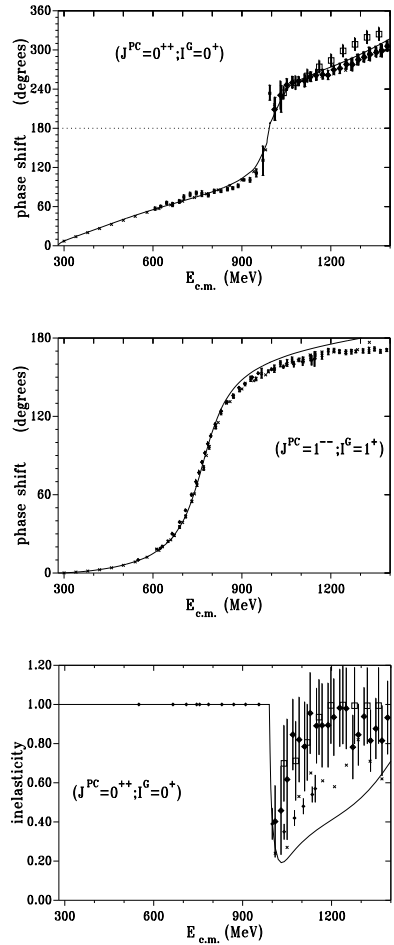} 
\caption{$\pi\pi$ phase shifts in the $J^P=0^+$ ($\sigma$) and $J^P=1^-$ 
	($\rho$) channels and the corresponding inelasticity in the $\sigma$ 
	channel.}
\label{fig8}
\end{figure}

The bare parameters (masses, coupling constants) in the pole diagrams
are dressed by unitarizing the interaction terms in a relativistic
Schr\"odinger equation.
The $K\anti{K} \rightarrow K\anti{K}$ interaction (Fig.~\ref{fig6},
third diagram) is strongly isospin dependent: in the scalar-isoscalar
channel all contributions ($\rho$, $\omega$, $\phi$) add up
and provide a sizable attraction, which leads to a $K\anti{K}$ bound
state at $f_0(975)$ (see Fig.~\ref{fig8} for the resulting phase shifts)
--- the genuine scalar resonance $f_0(1440)$ sits at a higher energy,
at about 1.4 GeV.
On the other hand, in the vector-isovector channel there is strong
cancellation between $\rho$ and $\omega$, $\phi$ exchange, since the
former changes sign.
Consequently, the influence of the $K\anti{K}$ channel here is
negligible.
The corresponding phase shift is dominated by the $\rho$-pole diagram,
as illustrated in Fig.~\ref{fig8}.

\subsection{$B\anti{B} \rightarrow 2\pi $ Helicity Amplitudes} 

Based on the $\pi\pi \rightarrow \pi\pi$ amplitude (which has a well
defined off-shell behavior) the evaluation of diagrams such as in
Fig.~\ref{fig1}(c) for any $BB'$ system can be done in two steps.
Firstly the $N\anti{N} \ (\Lambda \anti{\Lambda}, \ \Sigma \anti{\Sigma},
\ {\rm etc}.) \rightarrow 2\pi$ amplitudes (illustrated in Fig.~\ref{fig5})
are determined in the pseudophysical region ($t \leq 4 m^2_\pi$). 
For the transition Born amplitude, $V$, both the $N$ and $\Delta$
(or $\Lambda$, $\Sigma$ and $Y^*$ in case of $Y\anti{Y'}$) exchanges are 
taken into account.
Corresponding coupling constants in the transition interaction are
taken from the Bonn $NN$ \cite{MHE} and the J\"ulich $YN$ potential
models \cite{Holz}.
Note that this cannot be done for the cutoff masses at the vertices
since the form factors now act in quite a different kinematic regime,
where the baryon is the essential off-shell particle.
For the $NN$ case, the parameters can be fixed independently
(to $\Lambda_{NN\pi} = 1.9$ GeV, $\Lambda_{N\Delta \pi} = 2.1$ GeV)
by using quasi-empirical information obtained by 
H\"ohler at el.~\cite{Hoehler2} by analytically continuing the $\pi N$ 
and $\pi \pi$ scattering data.
Such information is, however, not available for the $Y\anti{Y}$
($Y = \Lambda, \Sigma$) channels, so that here we make the reasonable
assumption that
$\Lambda_{\Lambda\Sigma\pi} \simeq \Lambda_{\Sigma\Sigma\pi} \simeq 
\Lambda_{NN\pi}$
and $\Lambda_{\Lambda Y^*\pi} \simeq \Lambda_{\Sigma Y^* \pi} \simeq
\Lambda_{N\Delta\pi}$.

\section{POTENTIAL FROM CORRELATED $\pi\pi$ AND $K\anti{K}$ EXCHANGE}

{}From the $B\anti{B'} \rightarrow 2\pi $ helicity amplitudes the
spectral functions can be calculated (see Ref.~\cite{REUBER} for
details), which are then inserted into dispersion integrals to
obtain the (on-shell) baryon-baryon interaction:
\begin{equation}
V^{(0^+,1^-)}_{B_1',B_2';B_1,B_2}(t) \propto \int_{4m^2_\pi}^\infty
dt' 
{\rho^{(0^+,1^-)}_{B_1',B_2';B_1,B_2}(t') \over t'-t}, \ \  t < 0 .
\end{equation}

We should note that the helicity amplitudes obtained according to
Fig.~\ref{fig5} still generate the uncorrelated (first diagram on
the r.h.s. of Fig.~\ref{fig5}), as well as the correlated pieces
(second and third diagrams).
Thus, in order to obtain the contribution of the truely correlated
$\pi\pi$ and $K\anti{K}$ exchange we must eliminate the former from
the spectral function.
This can be done by calculating the spectral function generated by
the Born term and subtracting it from the total spectral function:
\begin{equation}
\rho^{(0^+,1^-)} \longrightarrow \rho^{(0^+,1^-)} - 
\rho^{(0^+,1^-)}_{\rm Born} .
\end{equation}

Note that the spectral functions characterize both the strength
and range of the interaction.
Clearly, for sharp mass exchanges the spectral function becomes
a $\delta$-function at the appropriate mass.

It is convenient to present our results in terms of effective
coupling strengths, by parameterizing the correlated processes
by (sharp mass) $\sigma$ and $\rho$ exchanges.
The interaction potential resulting from the exchange of a
$\sigma$ meson with mass $m_\sigma$ between two $J^P=1/2^+$
baryons $A$ and $B$ has the structure:
\begin{equation}
V^{\sigma}_{A,B;A,B}(t) \ = \ g_{AA\sigma} g_{BB\sigma} 
{F^2_\sigma (t) \over t - m^2_\sigma} , 
\end{equation}
where a form factor $F_\sigma(t)$ is applied at each vertex,
taking into account the fact that the exchanged $\sigma$ meson is
not on its mass shell. 
This form factor is parameterized in the conventional monopole form, 
\begin{equation}
F_\sigma (t) = {\Lambda ^2_\sigma - m^2_\sigma \over 
\Lambda ^2_\sigma - t} \ , 
\end{equation}
with a cutoff mass $\Lambda_\sigma$ assumed to be the same
for both vertices.
The correlated potential as given in Eq.~(1) can now be
parameterized in terms of $t$-dependent strength functions
$G_{B_1',B_2';B_1,B_2}(t)$, so that for the $\sigma$ case:
\begin{equation}
V^{(0^+)}_{A,A;B,B}(t) = 
G^{\sigma}_{A,A;B,B}(t) {1 \over t - m^2_\sigma}. 
\end{equation}
The effective coupling constants are then defined as:
\begin{equation}
g_{AA\sigma}g_{BB\sigma} \quad\longrightarrow \quad G_{AB\to
AB}^\sigma (t)= {(t-m_\sigma^2)\over\pi F^2_\sigma(t)}
\int_{4m_\pi^2}^{\infty} { \rho^{(0+)}_{AB;AB}(t') \over t'-t} dt' .
\label{eq:3_effccsig}
\end{equation}

Similar relations can be also derived for the correlated exchange
in the isovector-vector channel \cite{REUBER}, which in this case
will involve vector as well as tensor coupling pieces.

\vfill \eject

We stress once more that this parameterization does not involve any
approximations as long as the full $t$-dependence of the effective
coupling strengths is taken into account.
The parameters of $\sigma$ and $\rho$ exchange are chosen to have
the same values in all particle channels.
The masses $m_\sigma$ and $m_\rho$ of the exchanged particles have
been set to the values used in the Bonn-J\"ulich models of the
$NN$~\cite{MHE} and $YN$~\cite{Holz} interactions,\ 
$m_\sigma=550$ MeV, $m_\rho=770$ MeV.
The cutoff masses $\Lambda_{\sigma}$ and $\Lambda_{\rho}$ have been
chosen so that the coupling strengths in the $S=0, -1$ baryon-baryon
channels vary only weakly with $t$.
The resulting values ($\Lambda_\sigma=2.8$ GeV, $\Lambda_\rho=2.5$ GeV)
are quite large compared to the values of the phenomenological
parameterizations used in Refs.~\cite{Holz,MHE}, and thus represent
rather hard form factors.

Note that in the OBE framework the three reactions $NN\rightarrow NN$,
$YN\rightarrow YN$, $YY \rightarrow YY$
are determined by two parameters (coupling constants) $g_{NN\sigma}$
and $g_{YY\sigma}$, whereas the correlated exchanges are
characterized by three independent strength functions, so that vertex
coupling constants cannot be determined uniquely.

In the physical region the strength of the contributions is to a large
extent governed by the value of $G$ at $t=0$. The values for the various
channels are shown in Table \ref{tab:5_6_10}.

Apart from the values for our full model, Table 1 contains results
obtained when uncorrelated contributions involving spin-1/2 baryons only
are subtracted from the spectral function of the invariant baryon-baryon
amplitudes. These are the proper values to be used for constructing
a $NN$ or $YN$ model based on simple OBE-exchange diagrams. 
For the full Bonn $NN$ model, contributions involving spin-3/2 baryons
also need to be subtracted, since the corresponding contributions are already 
treated explicitly in this model, namely via box diagrams with intermediate
$\Delta$-states as shown in Fig.~\ref{fig1}(a). 
Obviously processes involving spin-3/2 baryons increase the
correlated contribution, in practice by about 30\% in all channels.

Comparing the relative strengths of effective $\sigma$ exchange
in the various baryon-baryon channels,
one observes that the scalar-isoscalar part of correlated $\pi\pi$
and $K\anti{K}$ exchange in the $NN$ channel is about twice as
large as in both $YN$ channels, and 3 -- 4 times larger than in the
$S=-2$ channels.

\renewcommand{\arraystretch}{1.3} 
\begin{table}[h]
\[
\begin{array}{|r|r|rr|rrr|}
\hline
\multicolumn{7}{|c|}{G^\sigma_{AB\to AB}/4\pi } \\
\hline
&NN&\Lambda N&\Sigma N&\Lambda\Lambda&\Sigma\Sigma&\Xi N	\\
\hline
\mbox{full model}			&5.87	&2.82	&2.58	&1.52
	&1.72	&1.19	\\
\mbox{subtractions for OBE model}	&7.77	&3.81	&3.15	&2.00
	&2.31	&1.52	\\
\hline
\end{array}
\]
\caption{Effective $\sigma$ coupling strengths $G^\sigma_{AB\to AB}(t=0)$
	for correlated  $\pi\pi$ and $K\anti{K}$ exchange in the various
	baryon-baryon channels. (The meaning of the rows is given in the
	text.)}
\label{tab:5_6_10}
\end{table}
\renewcommand{\arraystretch}{1.1} 
 
\begin{figure}[ht]
\vskip 5cm 
\includegraphics{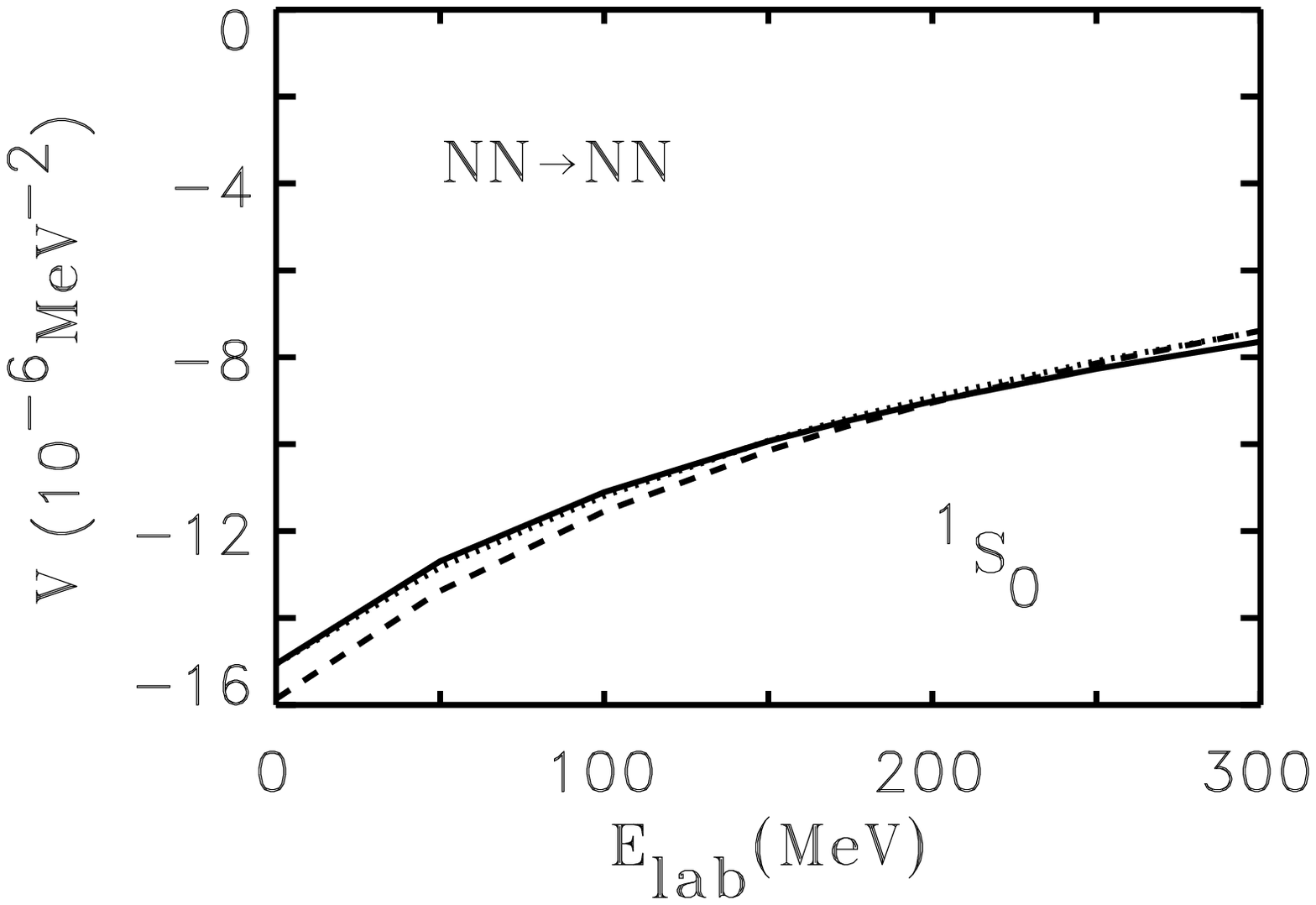} 
\includegraphics{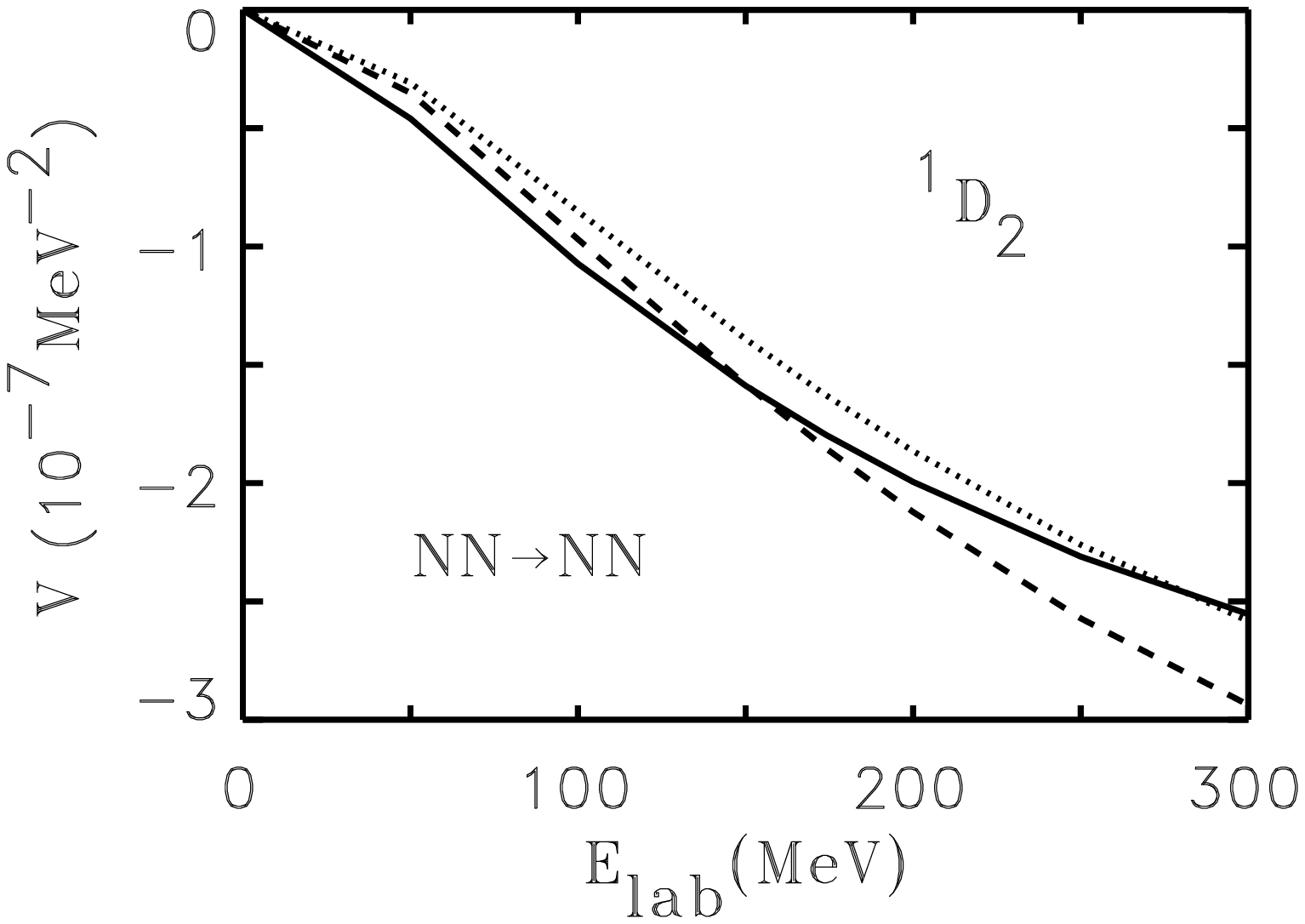}
\vskip 5cm 
\includegraphics{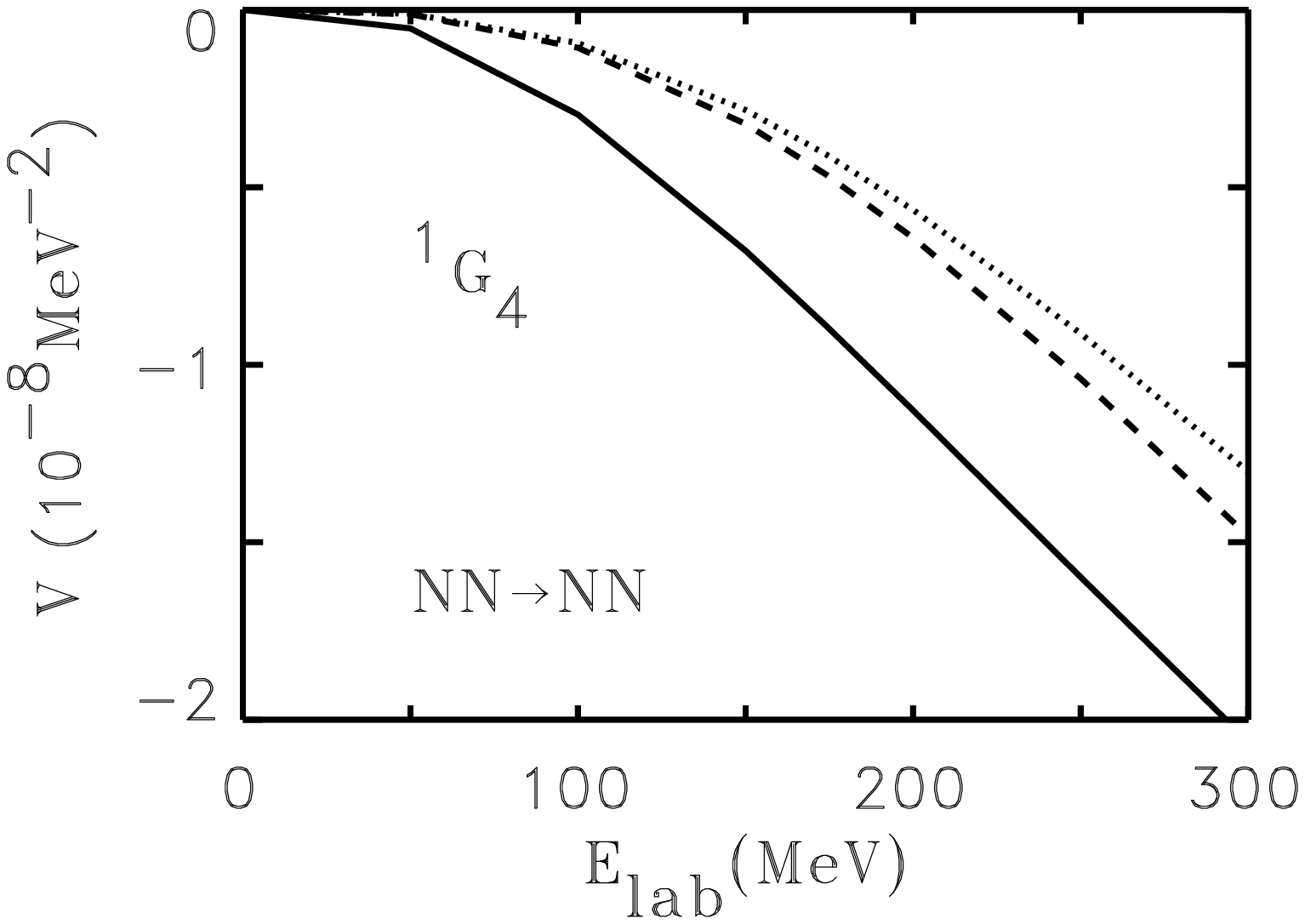}
\caption{{The $\sigma$-like part of the $NN$ on-shell potential in
various partial waves. The solid lines are derived from our microscopic
model of correlated $\pi\pi$ and $K\anti{K}$ exchange. The dotted lines
are obtained if the dispersion-theoretic result is parameterized by
$\sigma$ exchange, while the dashed correspond to the $\sigma$
exchange used in the Bonn OBEPT potential \protect\cite{MHE}. }} 
\label{fig:5_6_3}
\end{figure}

The average size of the effective coupling strengths is only an
approximate measure of the strength of correlated $\pi\pi$ and
$K\anti{K}$ exchange in the various particle channels.
The precise energy dependence of the correlated exchange as well
as its relative strength in the different partial waves of the
$s$-channel reaction is determined by the spectrum of exchanged
invariant masses, or spectral functions, leading to a different
$t$-dependence of the effective coupling strengths.
To demonstrate this we show in Fig.~\ref{fig:5_6_3} the on-shell
$NN$ potentials in spin-singlet states with angular momentum
$L=0, 2$ and 4, which are generated directly by the scalar-isoscalar
part of the correlated $\pi\pi$ and $K\anti{K}$ exchange.
As expected it is attractive throughout.

Figure \ref{fig:5_6_3} shows that the results evaluated from the
microscopic correlated $2\pi$ exchange model (solid curves) 
are comparable to those of the $\sigma$ exchange in the Bonn OBEPT 
potential (dashed curves) in partial waves with $J \leq 2$.  However, 
the correlated $2\pi$ exchange is significantly stronger in high partial 
waves because 
the $\sigma$ exchange, which corresponds to a spectral function proportional 
to $\delta(t'-m^2_\sigma)$, does not contain the long-range part of the 
correlated processes.  Indeed, parameterizing the results derived from 
the microscopic model by $\sigma$ exchange as before, but using the effective
coupling strength $G^\sigma_{NN\to NN}$ at $t=0$ (dotted curves), we 
obtain rough agreement with the exact result in the $^1S_0$ partial wave,
but underestimate the magnitude considerably in the high partial waves.
Obviously the replacement of correlated $\pi\pi$ and $K\anti{K}$
exchanges by an exchange of a sharp mass $\sigma$ meson with a
$t$-independent coupling cannot provide a simultaneous description
of both low and high partial waves.

Note that the above curves are based on the spectral function with 
subtraction for OBE models.
Corresponding results for the full model were presented in Fig.~22
of Ref.~\cite{REUBER}, and differ quantitatively from the ones shown
here.
In particular, the contribution of the correlated $\pi\pi$ and
$K\anti{K}$ exchanges in the scalar-isoscalar channel is stronger
than the $\sigma '$ exchange of the full Bonn potential in {\it all}
partial waves --- in agreement with earlier results reported in
Ref.\cite{Kim}.

\begin{figure}[ht]
\vskip 5cm 
\includegraphics{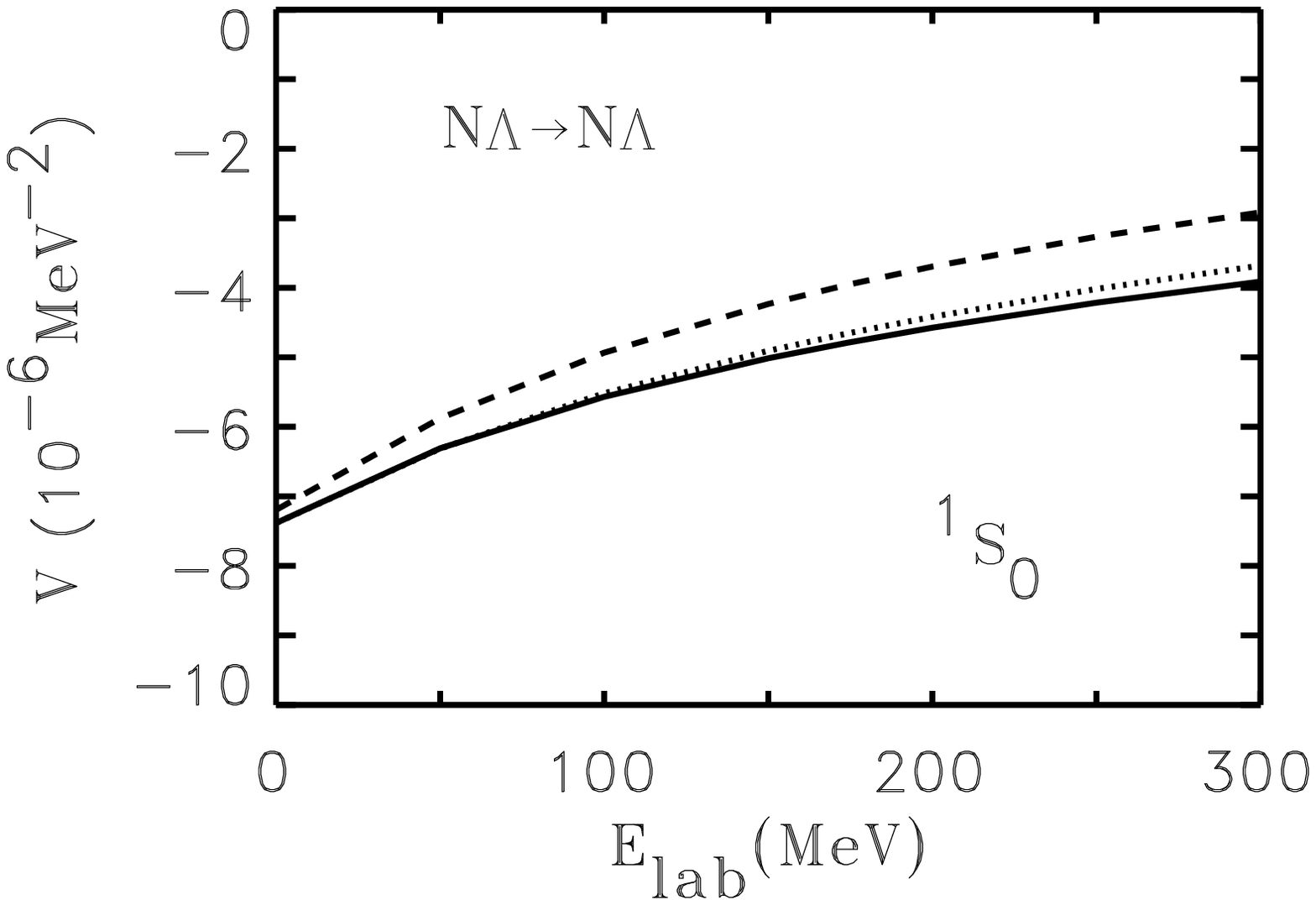} 
\includegraphics{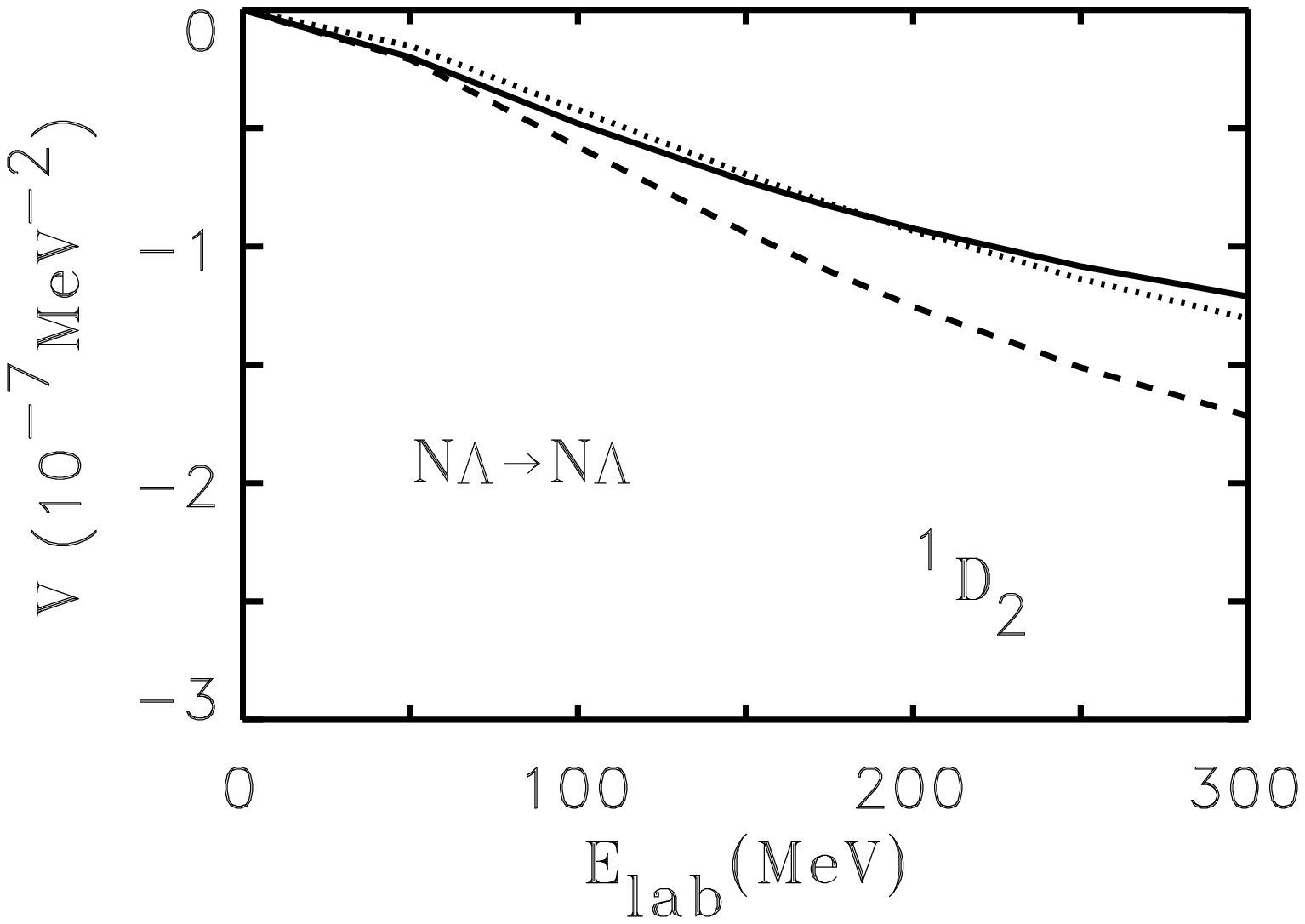} 
\vskip 5cm 
\includegraphics{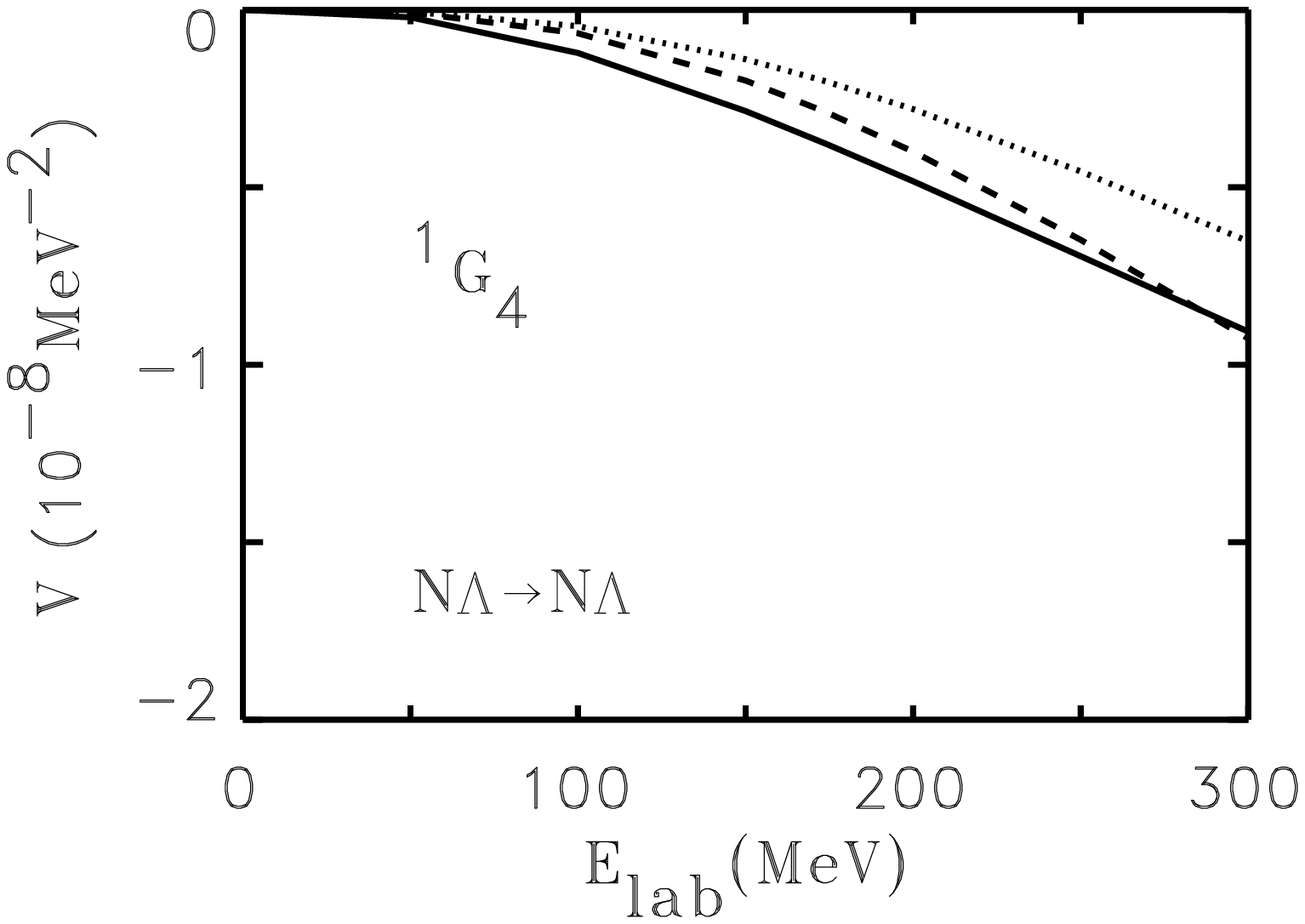} 
\caption{{The $\sigma$-like part of the $\Lambda N$ on-shell potential in
various partial waves. The curves are as in 
Fig.~\protect\ref{fig:5_6_3}, 
except for the dashed lines which 
correspond to the $\sigma$ exchange used in the J\"ulich $YN$ potential $A$
\protect\cite{Holz}. 
}} 
\label{fig:5_6_4}
\end{figure}

In Fig.~\ref{fig:5_6_4} we show the corresponding on-shell matrix
elements for the $\Lambda N$ channel.
Also here we see that the results generated by the scalar-isoscalar
part of correlated $\pi\pi$ and $K\anti{K}$ exchange are comparable
to the ones of the $\sigma$ exchange used in the J\"ulich $YN$ model~A.
In fact, correlated $\pi\pi$ exchange is slightly stronger in the
$^1S_0$ partial wave, but weaker in the $^1D_2$ partial wave.
Once again we see that the parameterization by an effective $\sigma$
exchange works well for the lower partial waves ($J\leq 2$), but fails
for the higher partial waves.

\begin{figure}[ht]
\vskip 5cm 
\includegraphics{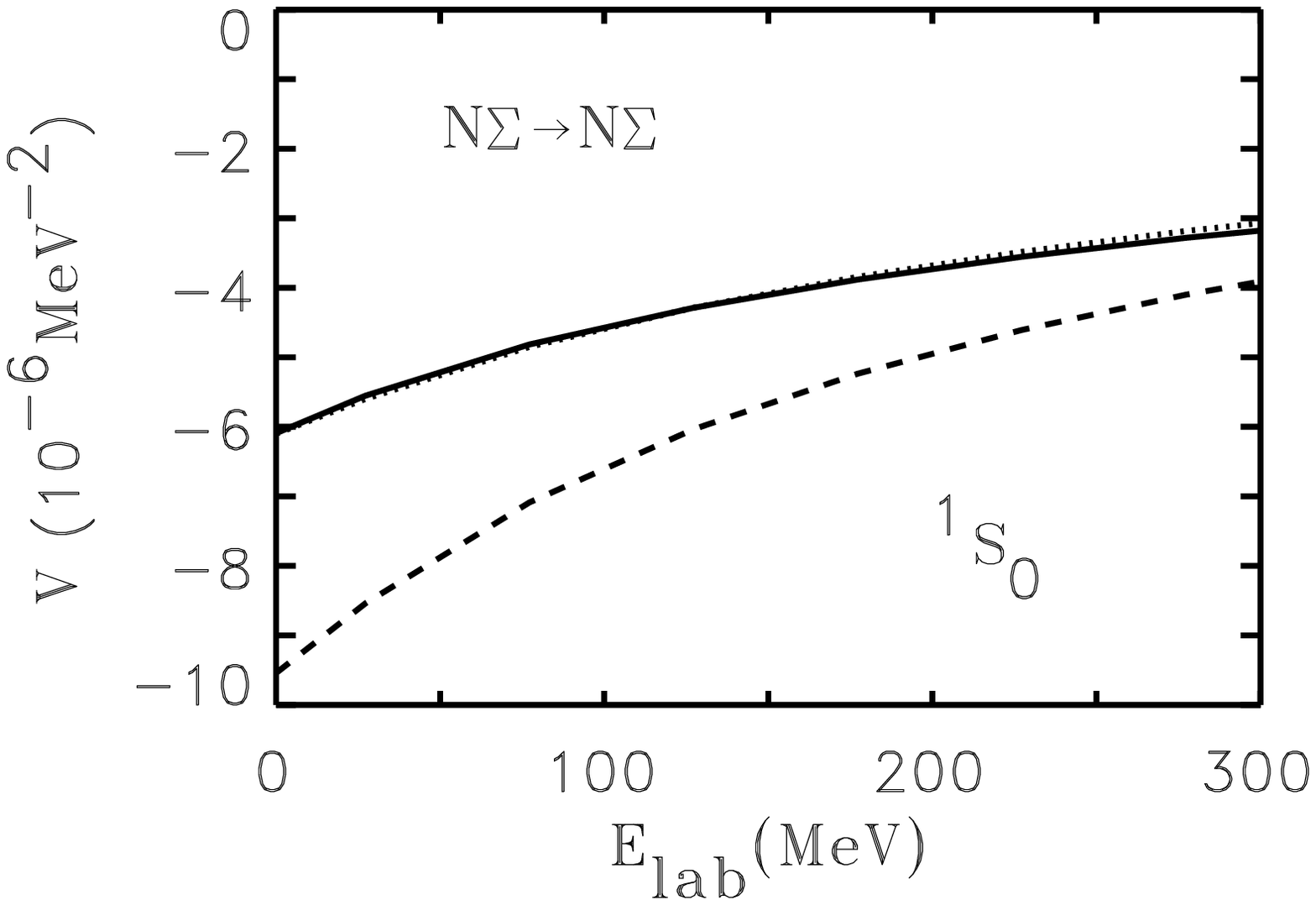} 
\includegraphics{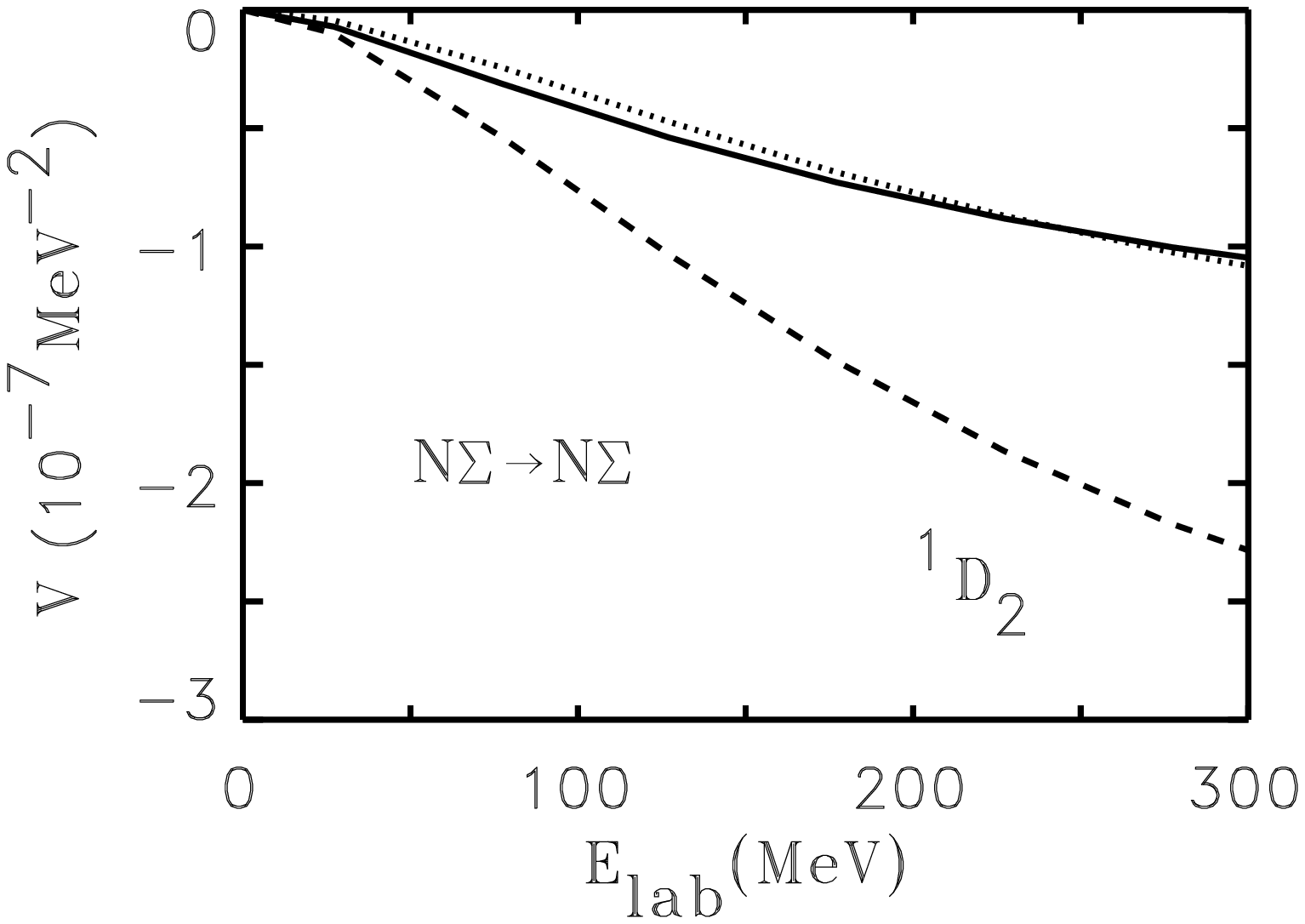} 
\vskip 5cm 
\includegraphics{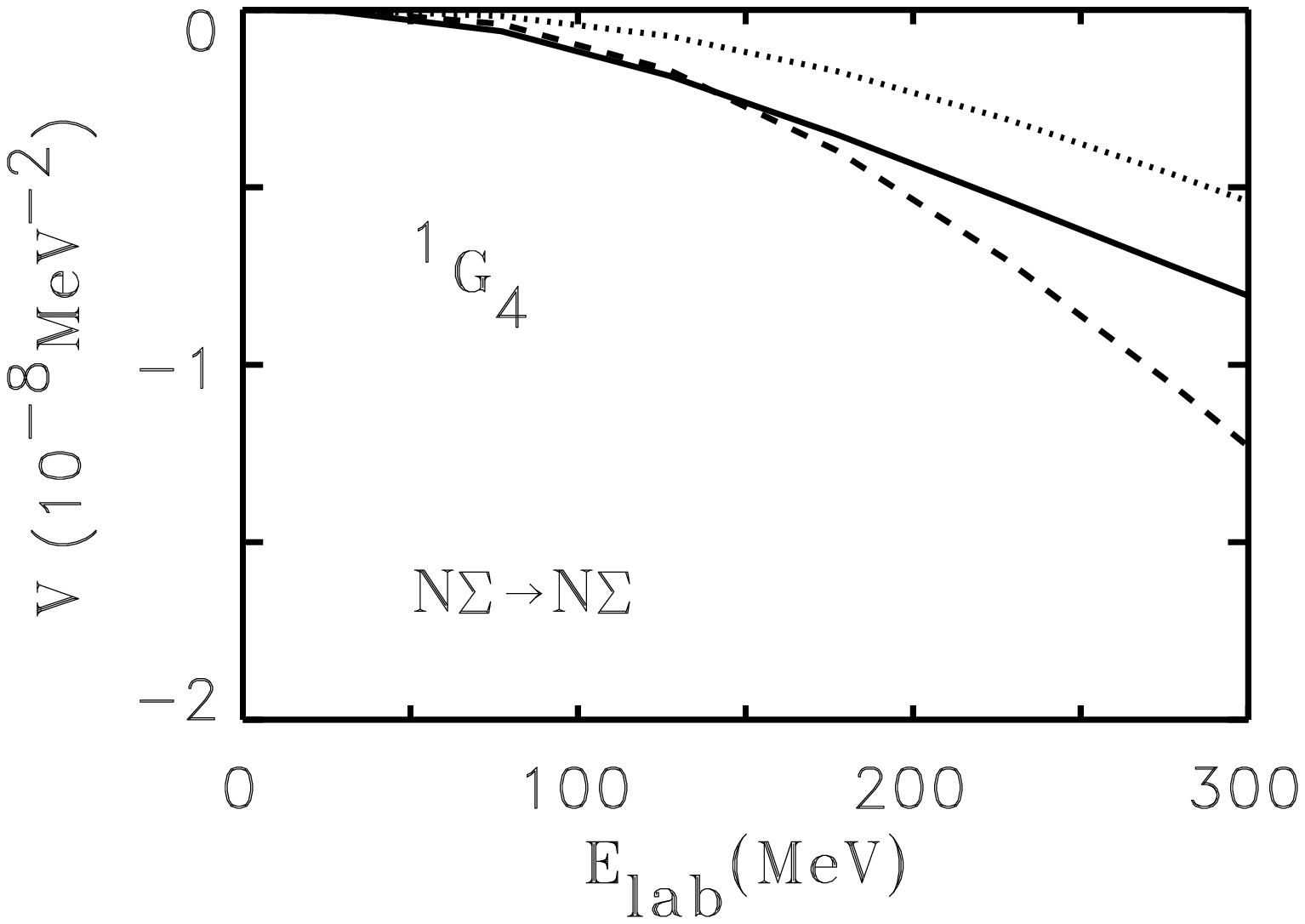} 
\caption{{The $\sigma$-like part of the $\Sigma N$ on-shell potential in
various partial waves. Same description of curves as in 
Fig.~\protect\ref{fig:5_6_4}. 
}} 
\label{fig:5_6_5}
\end{figure}

Finally, Fig.~\ref{fig:5_6_5} shows the corresponding results
for the on-shell $\Sigma N$ potentials.
Here one can see that the $\sigma$ exchange used in the J\"ulich $YN$
model A is clearly much stronger than the results one obtains from
the correlated $\pi\pi$ and $K\anti{K}$ exchange. 
These differences will have an impact on the properties of the new
hyperon-nucleon interaction model currently being developed \cite{we}.
Specifically, the weaker interaction in the scalar-isoscalar channel
resulting from our microscopic model of correlated $\pi\pi$ exchange 
provides much less attraction in the $N\Sigma$ $S$-waves and, in turn,
should also reduce the coupling between the $N\Lambda$ and $N\Sigma$
channels.
The strong coupling between these two systems in the original
J\"ulich $YN$ model is possibly one of the reasons why it
does not lead to a bound state for the hypertriton \cite{Miya}.

\section{SUMMARY}

An essential part of baryon-baryon interactions is the strong
medium-range attraction, which in one-boson-exchange models is
parameterized by exchange of a fictitious scalar-isoscalar meson
with mass around 500 MeV.
In extended meson exchange models this part is naturally generated
by two-pion exchange contributions.
As well as uncorrelated two-pion exchange, correlated contributions
must be included in which the exchanged pions interact during their
exchange; these terms in fact provide the main contribution to the
intermediate-range interaction.

In the scalar-isoscalar channel of the $\pi\pi$ interaction the coupling
to the $K\anti{K}$ channel plays a strong role, which has to be explicitly
included in any realistic model for energies near and above the $K\anti{K}$
threshold.
As kaon exchange is an essential part of hyperon-nucleon interactions
a simultaneous investigation of correlated $\pi\pi$ and $K\anti{K}$
exchanges is clearly necessary.
In Ref.~\cite{REUBER} the correlated $\pi\pi$ and $K\anti{K}$ exchange
contributions in various baryon-baryon channels have therefore been
investigated within a microscopic model for the transition amplitudes
of the baryon-antibaryon system ($B\anti{B'}$) into $\pi\pi$ and
$K\anti{K}$ for energies below the $B\anti{B'}$ threshold.
The correlations between the two mesons have been taken into account
by means of $\pi\pi-K\anti{K}$ amplitudes, determined in the field
theoretic framework of Refs.~\cite{Lohse,Janssen}, which provide an
excellent description of empirical $\pi\pi$ data up to 1.3 GeV.
With the help of unitarity and dispersion-theoretic methods, the
baryon-baryon amplitudes for correlated $\pi\pi$ and $K\anti{K}$
exchange in the $J^P=0^+$ ($\sigma$) and $J^P=1^-$ ($\rho$)
$t$-channels have then been determined.

In the $\sigma$ channel the strength of correlated $\pi\pi$ and
$K\anti{K}$ exchange decreases with the strangeness of the
baryon-baryon channels, becoming more negative.
In the $NN$ channel the scalar-isoscalar part of correlated
exchanges is stronger by about a factor of 2 than in both
hyperon-nucleon channels ($\Lambda N$, $\Sigma N$), and 3 to 4 
times stronger than in the $S=-2$ channels ($\Lambda \Lambda$,
$\Sigma\Sigma$, $N\Xi$).

With the current model it is now possible to reliably take into
account correlated $\pi\pi$ and $K\anti{K}$ exchange in both the
$\sigma$ and $\rho$ channels for various baryon-baryon reactions.
This will be especially important in processes where only scant
empirical data exist, as the elimination of the phenomenological
$\sigma$ and $\rho$ exchanges considerably enhances the predictive
power of baryon-baryon interaction models.
Given the strong constraints on $\sigma$ as well as $\rho$ exchange
from correlated $\pi\pi$ exchange, a more sound microscopic model
for the $YN$ interaction can hence now be constructed, which
can be used to address open questions such as the role of SU(3)
flavor symmetry, or the nature of the short range part of the $BB'$
interaction arising from $\omega$ exchange.

\bigskip 

{\bf Acknowledgements} 

\medskip

We would like to thank the Special Research Centre for the 
Subatomic Structure of Matter at the University of Adelaide
for support during the completion of this work.
J.H. was supported by the DFG project no. 477 AUS-113/3/0.

\def\Nucl{Nucl.\ }
\def\Phys{Phys.\ }
\def\Rev{Rev.\ }
\def\Lett{Lett.\ }
\def\PL{\Phys\Lett}
\def\PLB{\Phys\Lett B}
\def\NP{\Nucl\Phys}
\def\NPA{\Nucl\Phys A}
\def\NPB{\Nucl\Phys B}
\def\NPBS{\Nucl\Phys (Proc.\ Suppl.\ )B}
\def\PR{\Phys\Rev}
\def\PRL{\Phys\Rev\Lett}
\def\PRC{\Phys\Rev C}
\def\PRD{\Phys\Rev D}
\def\RMP{\Rev  Mod.\ \Phys}
\def\ZP{Z.\ \Phys}
\def\ZPA{Z.\ \Phys A}
\def\ZPC{Z.\ \Phys C}
\def\AOP{Ann.\ \Phys}
\def\PRep{\Phys Rep.\ }
\def\ANP{Adv.\ in \Nucl\Phys Vol.\ }
\def\PTP{Prog.\ Theor.\ \Phys}
\def\PTPS{Prog.\ Theor.\ \Phys Suppl.\ }
\def\PL{\Phys \Lett}
\def\JPF{J.\ Physique}
\def\FBSS{Few--Body Systems, Suppl.\ }
\def\IJMP{Int.\ J.\ Mod.\ \Phys A}
\def\NuCi{Nuovo Cimento~}

\end{document}